\documentclass[aps,prl,preprint,showpacs,superscriptaddress]{revtex4}
\usepackage[pdftex]{graphicx}
\usepackage{CJK}
\usepackage{xspace}
\usepackage[T1]{fontenc}
\usepackage{mathptmx}
\usepackage{multirow}
\usepackage{dcolumn} %centers to the decimal point in tables usind ``d''
\usepackage{pstricks}
\usepackage{psfrag}
 
\usepackage{xspace}  
\newcommand{\ts}{\textsuperscript}

\newcommand{\gammaray}{\ensuremath{\gamma}-ray\xspace} 
\newcommand{\gammarays}{\ensuremath{\gamma}-rays\xspace}
\newcommand{\mevu}{MeV/nucleon\xspace}
\newcommand{\etwop}{\ensuremath{E(2^+_1)}\xspace}

\newcommand{\stwop}{\ensuremath{2^+_1}\xspace}

\newcommand{\etwopt}{\ensuremath{2^+_1 \rightarrow 0^{+}_{\mathrm{gs}}}\xspace}

\newcommand{\beupl}{\ensuremath{B(E2;~0^+_{\mathrm{g.s.}}\rightarrow2^+_1)}\xspace} % long B(E2)
\newcommand{\beup}{\ensuremath{{B(E2)\!\!\uparrow}}\xspace}  % short B(E2)
\newcommand{\bethree}{\ensuremath{{B(E3)\!\!\uparrow}}\xspace}  % short B(E3)
\newcommand{\eb}{\ensuremath{e^{2}}b\ensuremath{^{2}}\xspace}
\newcommand{\eebbb}{\ensuremath{e^{2}}b\ensuremath{^{3}}\xspace}
\newcommand{\betwosn}{0.163(26)\xspace}
  
\begin{document}
\begin{CJK}{UTF8}{min}
  
  %  
  % frontmatter
  % 
  \title{Intermediate-energy Coulomb excitation of \ts{104}Sn:\\ Moderate $E2$ strength decrease
  approaching \ts{100}Sn}

  % commands for affiliation
  \newcommand{\ariken}{   \affiliation{RIKEN Nishina Center, Wako, Saitama 351-0198, Japan}}
  \newcommand{\apeking}{  \affiliation{State Key Laboratory of Nuclear Physics and Technology, Peking University, Beijing 100871, P.R.China}}
  \newcommand{\atum}{     \affiliation{Physik Department E12, Technische Universit\"at M\"unchen, 85748 Garching, Germany}}
  \newcommand{\alpc}{     \affiliation{LPC-Caen, ENSICAEN, Universit\'e de Caen, CNRS/IN2P3, 14050 Caen cedex, France}}
  \newcommand{\arikkyo}{  \affiliation{Department of Physics, Rikkyo University, Toshima, Tokyo 172-8501, Japan}}
  \newcommand{\auot}{     \affiliation{Department of Physics, University of Tokyo, Bunkyo, Tokyo 113-0033, Japan}}
  \newcommand{\atit}{     \affiliation{Department of Physics, Tokyo Institute of Technology, Meguro, Tokyo 152-8551, Japan}}
  \newcommand{\acns}{     \affiliation{Center for Nuclear Study, The University of Tokyo, RIKEN Campus, Wako, Saitama 351-0198, Japan}}
  \newcommand{\atus}{     \affiliation{Department of Physics, Tokyo University of Science, Noda, Chiba 278-8510, Japan}}
  \newcommand{\asaitama}{ \affiliation{Department of Physics, Saitama University, Saitama 338-8570, Japan}}
  \newcommand{\agsi}{     \affiliation{GSI Helmholtzzentrum f\"ur Schwerionenforschung GmbH, 64291 Darmstadt, Germany}}
  \newcommand{\aatomki}{  \affiliation{Institute for Nuclear Research of the Hungarian Academy of Sciences,  Debrecen, H-4001, Hungary}}
  \newcommand{\acea}{     \affiliation{CEA Saclay, Service de Physique Nucl\'eaire, F-91191 Gif-sur-Yvette, France}}
  \newcommand{\acsic}{    \affiliation{Instituto de Estructura de la Materia, CSIC, E-28006 Madrid, Spain}}
  \newcommand{\acracow}{  \affiliation{The Niewodniczanski Institute of Nuclear Physics, Polish Academy of Sciences, 31-342 Krakow, Poland}}
  \newcommand{\arcnp}{    \affiliation{Research Center for Nuclear Physics, Osaka University, Ibaraki, Osaka 567-0047, Japan}}
  \newcommand{\ainct}{    \affiliation{Ichinoseki National College of Technology, Ichinoseki-shi 021-8511, Japan}}

  \newcommand{\aem}{\email{pieter@ribf.riken.jp}}  % author email

  \author{P.~Doornenbal}                    \aem \ariken          % EMAIL: pieter@ribf.riken.jp
  \author{S.~Takeuchi}           \ariken              % EMAIL: takesato@ribf.riken.jp
  % in-beam gamma team and vip
  \author{N.~Aoi}               \arcnp                % EMAIL: aoi@riken.jp
  \author{M.~Matsushita}       \acns \arikkyo       % EMAIL: matsushi@cns.s.u-tokyo.ac.jp
  \author{A.~Obertelli}                     \acea                 % EMAIL: alexandre.obertelli@cea.fr
  \author{D.~Steppenbeck}                    \acns                % EMAIL: steppenbeck@riken.jp
  \author{H.~Wang}                 \ariken \apeking     % EMAIL: wanghe@ribf.riken.jp 
  % - frome here in alphabetical order
  \author{L. Audirac}                       \acea                 % EMAIL: laurent.audirac@cea.fr
  \author{H.~Baba}            \ariken               % EMAIL: baba@ribf.riken.jp
  \author{P.~Bednarczyk}                    \acracow              % EMAIL: piotr.bednarczyk@ifj.edu.pl
  \author{S.~Boissinot}                     \acea                 % EMAIL: simon.boissinot@cea.fr
  \author{M.~Ciemala}                       \acracow              % EMAIL: michal.ciemala@ifj.edu.pl
  \author{A.~Corsi}                         \acea                 % EMAIL: acorsi@cea.fr
  \author{T.~Furumoto}         \ainct                % EMAIL: furumoto@ichinoseki.ac.jp
  \author{T.~Isobe}            \ariken               % EMAIL: isobe@riken.jp
  \author{A.~Jungclaus}                     \acsic                % EMAIL: andrea.jungclaus@iem.cfmac.csic.es
  \author{V.~Lapoux}                        \acea                 % EMAIL: valerie.lapoux@cea.fr
  \author{J.~Lee}                \ariken              % EMAIL: jennylee@ribf.riken.jp
  \author{K.~Matsui}          \auot                % EMAIL: matsui@nucl.phys.s.u-tokyo.ac.jp
  \author{T.~Motobayashi}       \ariken              % EMAIL: motobaya@riken.jp
  \author{D.~Nishimura}       \atus                % EMAIL: dnishimura@rs.tus.ac.jp
  \author{S.~Ota}             \acns                 % EMAIL: ota@cns.s.u-tokyo.ac.jp
  \author{E.C.~Pollacco}                      \acea                 % EMAIL: e.pollaco@cea.fr
  \author{H.~Sakurai}          \ariken \auot        % EMAIL: sakurai@ribf.riken.jp
  \author{C.~Santamaria}                    \acea                 % EMAIL: Clementine.SANTAMARIA@cea.fr
  \author{Y.~Shiga}          \arikkyo              % EMAIL: 11la013t@rikkyo.ac.jp
  \author{D.~Sohler}                        \aatomki             % EMAIL: sohler@namafia.atomki.hu
  \author{R.~Taniuchi}          \auot                % EMAIL: taniuchi@nucl.phys.s.u-tokyo.ac.jp

  \date{\today}
 

  %  
  % abstract
  %  
  \begin{abstract}
    The reduced transition probability \beup of the first excited 2\ts{+} state in the nucleus \ts{104}Sn was 
    measured via Coulomb excitation in inverse kinematics at intermediate energies. A value of \betwosn \eb was 
    extracted from the absolute cross-section on a Pb target, while the method itself was verified with
    the stable \ts{112}Sn isotope. Our result deviates significantly from the earlier reported value of 0.10(4) \eb and
    corresponds to a moderate decrease of excitation strength relative to the almost constant values
    observed in the proton-rich, even-$A$ \ts{106--114}Sn isotopes. Present state-of-the-art shell-model 
    predictions, which include proton and neutron excitations across the 
    $N=Z=50$ shell closures as well as standard polarization charges, underestimate the experimental 
    findings. 
  \end{abstract}
  
  \maketitle  
\end{CJK} 

%  
% Introduction 
% 
Across the Segr\'e chart of nuclei the tin isotopes take an eminent position. Besides containing the longest chain of 
isotopes in-between two doubly-magic nuclei, in this case \ts{100}Sn and \ts{132}Sn, accessible to nuclear 
structure research, the valley of stability against $\beta$-decay crosses this chain at mid-shell. This  
allows for systematic studies of basic nuclear properties from very proton-rich $N=Z$ to very neutron-rich 
nuclei. Of high interest in this context is the robustness of the proton $Z=50$ shell closure when the 
$N=50,82$ magic numbers are approached. Experimentally, the $Z=50$ correlated gap size can be inferred from 
mass measurements when data from  neighboring isotones is available. The magnitude of the proton gap is well 
known for neutron-rich nuclei beyond the end of the major 
shell and shows a maximum for \ts{132}Sn~\cite{wang:2012:CPC}. On the proton-rich side, however, experimental
information is more scarce and only indirect evidence for a good $Z=50$ shell closure exists, e.g., the large 
Gamov-Teller strength observed in the $\beta$ decay of \ts{100}Sn~\cite{hinke:2012:NATURE}. 

% Why BE2 
%Besides mass measurements, 
Complementary shell evolution probes can be obtained from the 
\etwopt transition energies, \etwop, and their reduced transition probabilities \beupl, in short \beup.  
While the \etwop values of the tin isotopes between \ts{100}Sn and \ts{132}Sn
are well established and exhibit only very little variation -- the highest value is 1.472 MeV for 
\ts{102}Sn, the lowest is 1.132 MeV for \ts{124}Sn~\cite{lipoglavsek:1996:ZPA,raman:2001:ADNDT} -- 
the \beup transition strengths follow a different pattern. The a priori expectation is a curve showing maximum
collectivity at mid-shell and smoothly decreasing towards the shell closures, reflecting the number 
of particles times the number of holes available within the major shell. This perception is put on a formal
base for a single $j$-shell by the seniority scheme~(see, e.g., Ref.~\cite{casten:2001}), which predicts
constant \etwop excitation energies and a parabolic pattern for the transition strengths.
%Generalized seniority scheme 
It has been shown that these key characteristics remain valid in the generalized seniority scheme as long as 
the orbits within the major shell are filled with the same rate, while for different
level occupancies a shallow minimum for the \beup values can be obtained at mid-shell~\cite{morales:2011:PLB}. 
  
% The experimental situation
In recent years, several experimental findings generated the large interest
regarding the $E2$ strengths pattern in the tin isotopes. 
While the neutron-rich isotopes with $A=126,128,130$ follow the anticipated trend of
smoothly decreasing \beup values towards the major shell closure ~\cite{radford:2004:NPA,allmond:2011:PRC}
well described by large-scale shell-model (LSSM) calculations~\cite{banu:2005:PRC,ekstrom:2008:PRL},
the proton-rich nuclei take a different path. Commencing with the stable $A=114$ isotope a steadily growing
deviation from the shell-model expectations was observed with almost 
constant \beup values for the $A=106-112$    
isotopes~\cite{banu:2005:PRC,cederkall:2007:PRL,vaman:2007:PRL,ekstrom:2008:PRL,
doornenbal:2008:PRC,kumar:2010:PRC}. 
This triggered a revisit of the stable tin isotopes via direct lifetime
measurements, yielding generally lower $E2$ transition strengths than the adopted values given in 
Ref.~\cite{raman:2001:ADNDT} and even a local minimum for the mid-shell $A=116$ nucleus~\cite{jungclaus:2011:PLB}. 
%This minimum and the enhanced \beup values for the proton-rich nuclei can be reproduced 
%by introducing different cores and neutron effective charges for the upper and lower half of the major shell~\cite{jiang:2012:PRC},
%whereas using isospin-dependent neutron effective charges produces a maximum around $N=58$ ~\cite{baeck:2013:PRC}.
    
While the \beup value of \ts{102}Sn remains missing for a complete pattern of the tin isotopes
within the major shell, a first attempt for \ts{104}Sn with limited statistics has recently been 
made~\cite{guastalla:2013:PRL}. The result of 0.10(4) \eb indicates a steep decrease of excitation 
strength in agreement with LSSM calculations. 
%However, the large peak width obtained for the calibration run on \ts{112}Sn 
%is inconsistent with an earlier work measuring the same nucleus with the same \gammaray detector 
%setup~\cite{banu:2005:PRC} and cannot be inferred from lifetime effects~\cite{doornenbal:2010:NIMA}. 
%Considering the low statistics and the bad signal to noise ratio, the increased
%width may originate from unaccounted background, thereby leading to a too low \beup value for \ts{104}Sn.
%However, proton-rich unstable tin beams, necessary for Coulomb excitation experiments in inverse kinematics, are a 
%challenge for current research facilities. Due to the ongoing progress and continuous developments of accelerators 
%and separation techniques, a total of 40 tin isotopes have been observed so far. While on the neutron-rich side 
%the combination of induced fission and in-flight separation have led to the discovery of the heaviest tin isotope 
%with mass $A=138$~\cite{ohnishi:2010:JPSJ}, well beyond the $N=82$ shell closure, only very few doubly-magic 
%$N=Z$ \ts{100}Sn isotopes have been produced so far in fragmentation experiments using \ts{112}Sn and \ts{124}Xe 
%primary beams~\cite{hinke:2012:NATURE,bazin:2008:PRL,lewitowicz:1994:PLB}. 
% What was done
% 104Sn, any proton-rich tin absolute cross-section, any exotic nuclei absolute cross-section
In order to ameliorate the experimental situation, a new measurement of the \stwop transition strength in
\ts{104}Sn is desirable. Here, we report on the first \beup extraction in the unstable, proton-rich tin
nuclei from absolute Coulomb excitation cross-sections. Previously deduced values relied on target excitation
at ``safe''~\cite{cederkall:2007:PRL,ekstrom:2008:PRL} and intermediate~\cite{vaman:2007:PRL} energies or 
used a stable tin isotope with known excitation strength as normalization~\cite{banu:2005:PRC,guastalla:2013:PRL}. 
In fact, all reported values from intermediate-energy Coulomb excitation measurements above 100~\mevu rely on 
the latter method~\cite{banu:2005:PRC,burger:2005:PLB,saito:2008:PLB} and so far no attempt has been made to 
determine absolute cross-sections at these high energies. Therefore, in the present work the 
stable \ts{112}Sn isotope, which has a known \beup value, was Coulomb excited as well
in order to validate the method.  
  
%
% Experiment
%
The experiment was performed at the Radioactive Isotope Beam Factory (RIBF), operated 
by the RIKEN Nishina Center and the Center for Nuclear Study of the University of Tokyo.
A \ts{124}Xe primary beam was accelerated up to
an energy of 345~\mevu and impinged on a 3~mm thick Be production target at the F0 focus of the 
BigRIPS fragment separator~\cite{kubo:2012:PTEP}. The $B\rho-\Delta E-B\rho$ 
method was applied to select and purify secondary beams of \ts{104}Sn and \ts{112}Sn in
two subsequent measurements. 
%Two wedge-shaped aluminum degraders of 3~mm thickness each were 
%inserted at the F1 and F5 dispersive focal points. 
%BigRIPS and Purification
The beam cocktail compositions were identified event-by-event.
An ionization chamber located at the focal point F7 measured the energy loss $\Delta E$,
yielding the fragments' element number $Z$. The combination of position and angle measurements at 
the achromatic focal point F3 and the dispersive focal point F5 with parallel plate avalanche 
counters (PPAC)~\cite{kumagai:2001:NIMA} and a time-of-flight (TOF) 
measurement with two plastic scintillators placed at the focal points F3 and F7 
enabled the deduction of the mass-to-charge ratio $A/Q$. For the \ts{104,112}Sn secondary beams,
momentum acceptances were 2.2\% and 0.9\%, respectively. 
%The secondary beam rates of the \ts{104,112}Sn fragments on the reaction target were 310 and 1400
%particles per second (pps), respectively, with beam purities of 23\% and 78\%. 
%Other main beam components were \ts{111}In for the \ts{112}Sn setting and for the \ts{104}Sn setting it was
%\ts{103}In and \ts{102}Cd.
   
% PID after secondary target  
%\begin{figure}  
%  \centering  
%  \includegraphics[width=8.3cm]{pidzds.pdf}
%  \caption{(color online) Particle identification plot behind the secondary target using
%    the ZDS beam line detectors. A gate was applied on incoming \ts{104}Sn particles.
%    Three different charge states are visible for the \ts{104}Sn ejectiles.} 
%  \label{fig:pid-zds}
%\end{figure}  

% Secondary target area  
The secondary beams were transported to the focal point F8, where a 557~mg/cm\ts{2} thick Pb target 
was inserted to induce Coulomb excitation reactions. At mid-target, the secondary beam energies were 
131 and 154~\mevu for the \ts{104,112}Sn fragments. In order to enhance the number of tin fragments
in the fully stripped charge state, a 6 mg/cm\ts{2} thick aluminum foil was placed behind the reaction target.
Scattering angles were determined with two PPACs located 1430 and 930~mm upstream  
and one PPAC located 890 mm downstream the secondary target. The PPACs' position 
resolution in X and Y was 0.5~mm ($\sigma$), allowing for a scattering angle reconstruction resolution of 
about 5~mrad, while an angular straggling of 6--8~mrad was calculated with the ATIMA code~\cite{atima}. 
Grazing angles, calculated using the formulas given in Ref.~\cite{wollersheim:2005:nima}, 
were 28 and 23~mrad for \ts{104,112}Sn and their respective energies in front of the reaction target. 
Due to the scattering angle resolution
and the angular straggling, a cut  on ``safe'' angles would have led to a  
loss of a large fraction of the \gammaray yield. Therefore, no angular cut was applied and contributions 
from nuclear excitations were determined from inelastic scattering on a 370~mg/cm\ts{2} thick carbon target.  

% Observed spectra
\begin{figure} 
  \centering
  \includegraphics[width=8.3cm]{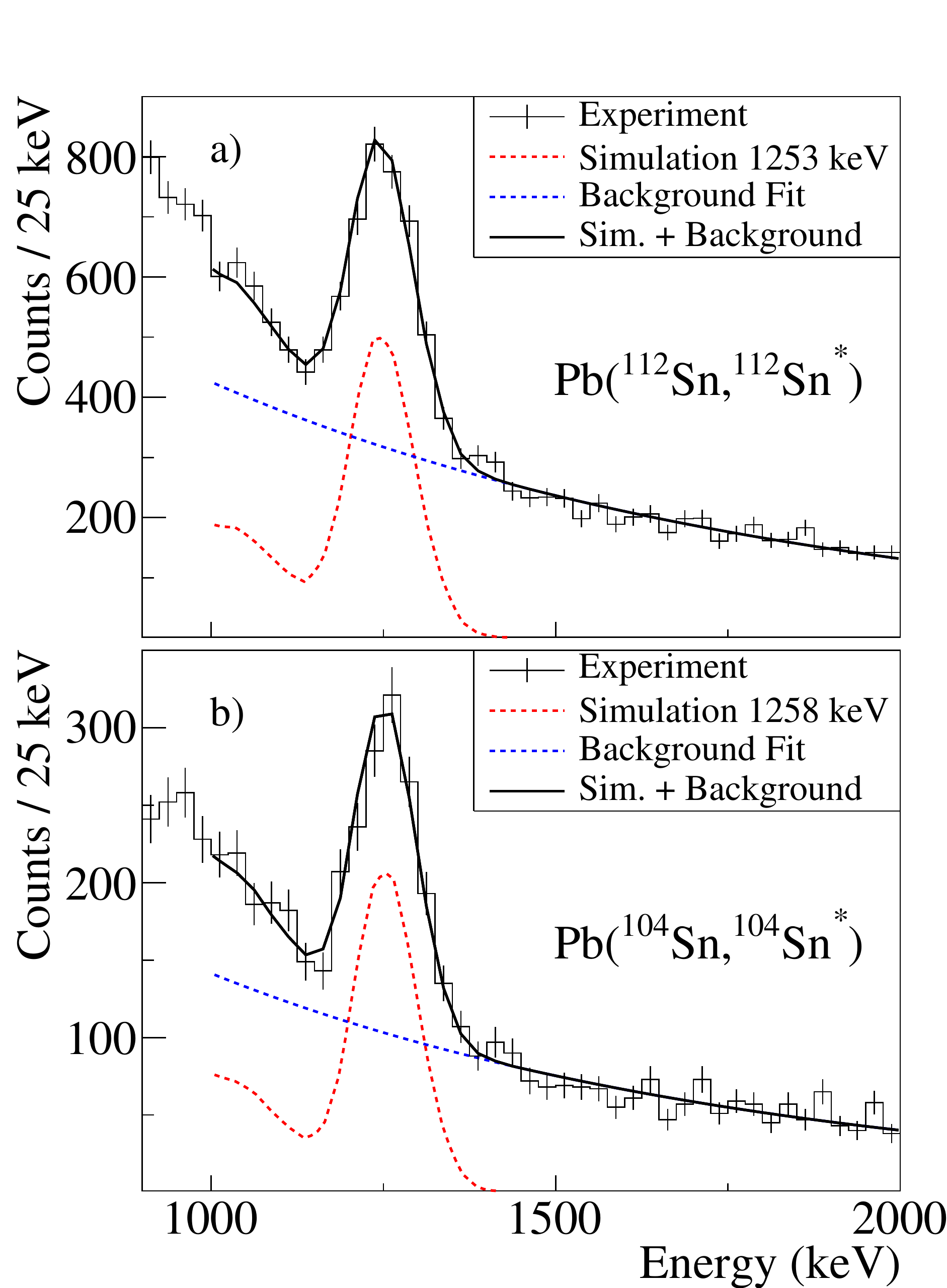}
  \caption{(color online)
    Doppler corrected \gammaray energy spectra following Coulomb excitation of
    fully-stripped \ts{112}Sn (a) and \ts{104}Sn (b) ejectiles detected in coincidence with the 
    BigRIPS and ZeroDegree spectrometers. Intensities were determined
    by fitting the observed line-shapes with simulated response functions (red dotted) on top of 
    exponential background (blue dotted). The resulting curves
    are shown by the solid lines.
  }
  \label{fig:spec}
\end{figure} 
  
%DALI2 and ZDS
To detect \gammarays from the \etwopt transitions, the reaction target was surrounded by the
DALI2 array~\cite{takeuchi:2003}. It consisted of 186 NaI(Tl) detectors, covering
center-of-crystal angles from 19 to 150~degrees. 
The efficiency of the DALI2 spectrometer was measured with \ts{60}Co and \ts{88}Y stationary sources
and agreed within 5~\% to simulations using GEANT4~\cite{agostinelli:2003:nima}.
For the 1.33~MeV \gammaray emitted by the stationary \ts{60}Co source, a full energy peak (FEP) 
detection efficiency of 14~\% and an energy resolution of 6~\% (FWHM) were measured for the full array.   
Radiation arising from secondary bremsstrahlung produced from the ions' deceleration in the reaction 
target was the anticipated main source of background. Therefore, the beam pipe at the F8 focus was 
enclosed by 1~mm of lead and 1~mm of tin shields. In addition, only forward angle detectors 
in the rest-frame were analyzed. After Doppler shift correction for a 1.26~MeV \gammaray emitted 
in-flight, values of 10~\% and 8~\% (FWHM) were expected for the FEP efficiency and energy resolution, 
respectively. %Add-back of \gammaray energies measured in more than one DALI2 detector was not applied.

% ZDS  
Reaction products behind the reaction target were identified by the ZeroDegree 
spectrometer~\cite{kubo:2012:PTEP}, using the previously described $\Delta E- B\rho-$TOF method 
from focus F8 to focus F11. 
%The $B\rho$ value was set to fully stripped \ts{104,112}Sn ejectiles and their position
%and angles measured by PPACs placed at the dispersive focus F9 and the achromatic focus F11.
%The velocity was measured by two plastic scintillators at the foci F8 and F11 and the ionization
%chamber was placed at F11. 
%The particle identification spectrum for the \ts{104}Sn setting
%is displayed in Fig.~\ref{fig:pid-zds}, which exhibits three charge states. 
Angular acceptances were $\pm 30$~mrad vertically and $\pm 45$~mrad horizontally for particles passing
ZeroDegree with the central momentum.
Including efficiencies of 83 and 76~\% for scattering angle determination, 180 and 920 particles per second
 of \ts{104,112}Sn ejectiles were detected in the ZeroDegree spectrometer in their fully-stripped charge state. 
%  
% Results
Figure~\ref{fig:spec} displays the \gammaray spectra measured
in coincidence with fully-stripped \ts{104,112}Sn ions detected in BigRIPS and ZeroDegree
after applying the Doppler shift correction. The two transitions were observed at 1258(6) and 1253(6)~keV, 
close to the literature values of 1260 and 1257~keV~\cite{raman:2001:ADNDT}.
%Besides a statistical error of 2 keV, errors of the peak position determination originated from the energy 
%calibration (3~keV), uncertainties of the average \gammaray emmission angles of the individual 
%detectors (3~keV), and beam velocities for the Doppler shift correction (3~keV). 
The intensities were determined by fitting the experimentally observed spectra 
%between 1000 and 2000 keV 
with simulated response functions on top of exponential background.

\begin{figure}  
  \centering
  \includegraphics[width=8.3cm]{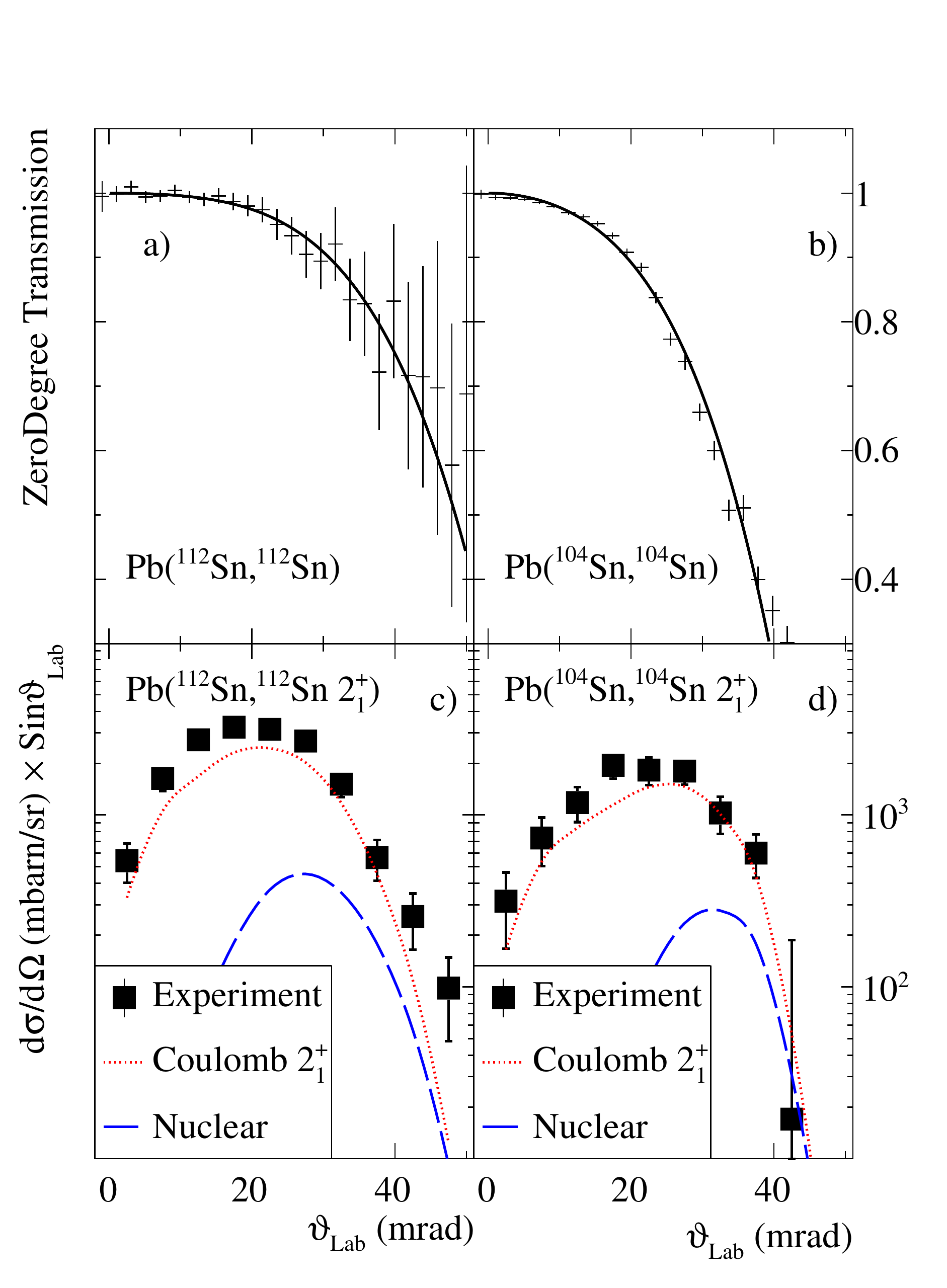}
  \caption{(color online) 
    The two top panels display the ZeroDegree transmission as function of the scattering angle 
    for \ts{112}Sn~(a) and \ts{104}Sn~(b) on the Pb target, in the two bottom panels the differential 
    inelastic scattering cross-sections are shown for 
    \ts{112}Sn~(c) and \ts{104}Sn~(d). The calculated distributions are Coulomb excitation of the \stwop state
    (red dotted) and the nuclear excitation (blue dashed). The difference between
    measured and the calculated cross-sections is attributed to Coulomb feeding from higher 
    lying states. See text for details.
    %Detector resolutions, angular straggling, and the scattering angle dependent ZeroDegree 
    %transmission were convoluted into the calculated curves. Larger errors in the transmission curve 
    %of \ts{112}Sn are obtained
    %due to a higher scale down factor for the beam particle trigger (100 compared to 5 for \ts{104}Sn).
  } 
  \label{fig:angular}
\end{figure}   

% Determination of the BE2
The measured inelastic cross-sections $\sigma_{2^+_1}$ are composed of contributions from nuclear
excitation, Coulomb excitation to the \stwop state, and Coulomb feeding from higher lying states 
($\sigma_{2^+_1} = \sigma_{n} + \sigma_{c} + \sigma_{f}$). In addition, the ZeroDegree
angular acceptance depends on the momentum distribution of the secondary beam and has to be corrected for. 
Thus, a measured cross-section on the Pb target can be converted to a \beup value only if
$\sigma_{n}$ and $\sigma_{f}$ are quantified. 

Inelastic scattering cross sections to the \stwop state of
 40(4) and 48(4) mbarn were measured for the \ts{104,112}Sn isotopes on the carbon target. These values were 
reproduced with the DWEIKO code~\cite{bertulani:2003:CPC} by selecting nuclear vibrational excitations and 
deformation lengths of 
$\delta = 0.38(2)$ and 0.45(2)~fm. Optical potentials were derived for the calculations as described in 
Ref.~\cite{furumoto:2012:PRC} using the microscopic folding model with the complex G-matrix interaction 
CEG07~\cite{furumoto:2008:PRC,furumoto:2009:PRC} and the density presented in Ref.~\cite{chamon:2002:PRC}. 
%$\beta_N = $ 0.097(4) and 0.089(4)
%The extracted
%deformation lengths correspond to $\beta_N = $ 0.097(4) and 0.089(4) for \ts{104,112}Sn, respectively,
%with 

For inelastic scattering of \ts{112}Sn on the Pb target, the cross-section to the \stwop state was $\sigma_{2^+_1}=524(39)$~mbarn,
while a cross-section of $\sigma_{n} + \sigma_{c} = 412(13)$ mbarn was calculated with DWEIKO using the \beup value of 
0.242(8)~\eb ~\cite{raman:2001:ADNDT,kumar:2010:PRC}, the derived deformation length, and the angular
transmission shown in Fig.~\ref{fig:angular}~(a). Thus, a Coulomb feeding contribution of $\sigma_{f}=112(41)$~mbarn 
was determined.  
Figure~\ref{fig:angular}~(c) displays the measured differential cross-section as function of 
scattering angle. It is compared to the calculated nuclear cross-sections and to the \stwop Coulomb excitation 
cross-section. Coulomb excitations are dominant for all scattering angles, as nuclear contributions are 
sizable only around the grazing angle. A larger nuclear contribution would have resulted in a 
maximum at the grazing angle. All calculations were convoluted with the detector resolution, the 
angular straggling, and the observed ZeroDegree scattering angle transmission. 
  
The Coulomb feeding can be attributed to single-step Coulomb excitations to higher lying states and subsequent 
decay via the \stwop state, while multi-step excitations play only a minor role at intermediate energies. 
For example, the \bethree value to the $3^-_1$ state at 2355~keV in \ts{112}Sn has a strength of 
0.087(12)~\eebbb \cite{jonsson:1981:NPA} and decays through the \stwop state. This translates to a feeding 
contribution of 10(2)~mbarn. For $E2$ excitations, the total strength measured in the 
heavier \ts{116--124}Sn isotopes between two and four MeV corresponds to about 10~\% of that to the first
excited state~\cite{bryssinck:2000:PRC}, while no experimental information is available for $2^+$ states
at higher energies. 
%At high beam 
%velocities, also the $E1$ 
%The exact 
%scattering angle strength distributions for the different multipolarities are impossible to calculate, as they 
%depend on the excitation energy of the individual levels. 

The observed Coulomb feeding contributions in \ts{112}Sn can be used to evaluate the Coulomb
feeding for \ts{104}Sn. In a simple picture, 
it originates from the $3^-_1$ excitation and fragmentation of the $E2$ excitation strength to many $2^+$ 
states between 2 MeV and the proton separation energy $S_p$ (7.554(5) MeV for \ts{112}Sn and 4.286(11) MeV 
for \ts{104}Sn~\cite{wang:2012:CPC}). Assuming an uniform $E2$ excitation strength distribution in this region, the same $E3$ 
excitation strength and correcting for the angular transmission shown in Fig.~\ref{fig:angular} (b) results in a 
Coulomb feeding of $\sigma_{f}= 46(19)$~mbarn for \ts{104}Sn, mainly due to the lower $S_p$ value. This estimation 
is corroborated by a higher peak-to-background ratio for \ts{104}Sn despite a lower total cross-section, showing that 
fewer high lying excited states are populated. 

\begin{figure} 
  \centering 
  \includegraphics[width=8.3cm]{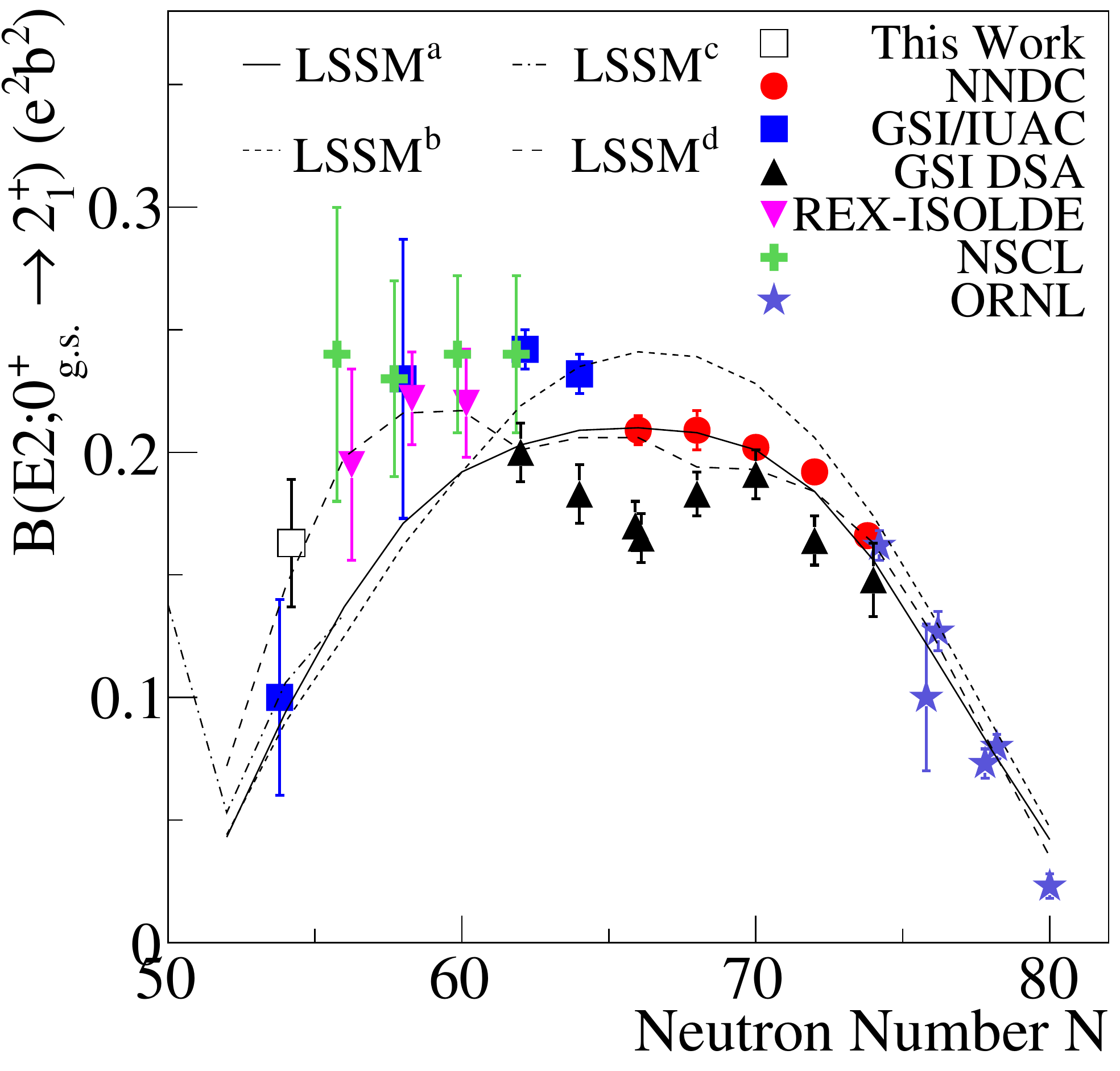}
  \caption{(color online) 
    Experimental \beup values for even-mass Sn isotopes in-between the doubly magic \ts{100}Sn 
    and \ts{132}Sn nuclei~\cite{banu:2005:PRC,cederkall:2007:PRL,vaman:2007:PRL,ekstrom:2008:PRL,doornenbal:2008:PRC,kumar:2010:PRC,radford:2004:NPA,allmond:2011:PRC,raman:2001:ADNDT}. Results from LSSM calculations 
    using different inert cores and effective charges are shown for comparison. See text for details. 
    %In the bottom panel, the proton gap size $\Delta_{\pi}$ was deduced from the atomic mass 
    %evaluation AME2012~\cite{wang:2012:CPC}.
  } 
  \label{fig:systematics}
\end{figure}   

%  
% Result for 104Sn and Discussion 
%  
For inelastic scattering of \ts{104}Sn on the Pb target, a cross section of $\sigma_{2^+_1}=298(30)$ mbarn was measured. Taking the 
previously determined nuclear contributions and the Coulomb feeding into account, a \beup of \betwosn \eb was deduced. 
Note that due to the lower beam energy and the reduced scattering angle acceptance, nuclear contributions were 
significantly suppressed ($\sigma_{n}=28(3)$~mbarn), as can be seen in the differential cross-section in Fig.~\ref{fig:angular}~(d). 
The new  \beup value is displayed in Fig.~\ref{fig:systematics} together with known data in-between the two doubly-magic 
tin nuclei. Our result deviates significantly from the value of 0.10(4)~\eb obtained in Ref.~\cite{guastalla:2013:PRL}. It 
corresponds to only about 30 \% excitation strength decrease compared to the even-$A$ \ts{106--114}Sn isotopes and
shows that the reduction is much more shallow than previously suggested.
   
Correlated with the recent experimental progress approaching \ts{100}Sn, various \beup calculations
have been presented~\cite{banu:2005:PRC,ekstrom:2008:PRL,morales:2011:PLB,jiang:2012:PRC,baeck:2013:PRC,guastalla:2013:PRL}.
Very instructive are the LSSM calculations presented with the first 
intermediate-energy Coulomb excitation experiment on proton-rich tin isotopes~\cite{banu:2005:PRC}.
Within that work, two sets of \beup calculations were performed, using \ts{100}Sn and \ts{90}Zr as
inert cores, respectively, and an effective interaction derived from the CD-Bonn potential~\cite{machleidt:2001:PRC}. 
The former, denoted LSSM\ts{a}, used a neutron model space with the 
$1d_{5/2}$, $0g_{7/2}$, $1d_{3/2}$, $2s_{1/2}$, and $0h_{11/2}$ orbitals, the latter, denoted LSSM\ts{b}, contained the
proton $0g_{9/2}$, $0_{g7/2}$, $1d_{5/2}$, $1d_{3/2}$, and $2s_{1/2}$ ($gds$) orbitals as well. 
Neutron effective charges of $e_{\nu} = 1.0e$ were used for the \ts{100}Sn core calculations to compensate the neglect of
proton excitations across the $Z=50$ shell  while the \ts{90}Zr core calculations allowed up to four-particle-four-hole proton 
excitations and used ``standard'' neutron and proton effective charges of $e_{\nu}=0.5e$ and $e_{\pi}=1.5e$. The results are added in 
Fig.~\ref{fig:systematics} to the experimental values between the two doubly magic nuclei and yield inverted parabola in agreement
with the neutron rich nuclei but fail to reproduce the \beup enhancement for proton-rich nuclei. 

In the most recent 
shell model prediction, denoted LSSM\ts{c} in Fig.~\ref{fig:systematics}, the calculations were expanded to a 
\ts{80}Zr core and a $gds$ model space, thereby allowing neutron as well as proton excitations across the $N=Z=50$ gap~\cite{guastalla:2013:PRL}.
The standard effective charges were used and truncation was applied depending on the 
nuclei's neutron number due to computational limits. However, the inclusion of neutron excitations across the $N=50$ gap 
augmented the \beup values only slightly. For \ts{104}Sn, a value of about 0.1~\eb is predicted, well below our experimental
finding, and also the experimental \beup values for \ts{106}Sn are underestimated.  

Different suggestions to break the symmetry in the theoretical \beup pattern have been made ranging from refined tuning of the proton-neutron 
monopoles~\cite{banu:2005:PRC,cederkall:2007:PRL}, inclusion of excitation across the $N=50$ shell~\cite{banu:2005:PRC}, a $N=50$ shell gap 
reduction~\cite{ekstrom:2008:PRL}, to simply using two different sets of single particle levels and effective charges for the lower and 
upper half of the shell~\cite{jiang:2012:PRC}. 
%In the latter work, valence neutrons in the lower half with mass $A<116$ 
%were treated as particles of a \ts{100}Sn inert core and the upper half with mass $A\ge116$ as holes in a \ts{132}Sn inert core. 
%A large effective charge of $e_{\nu} = 1.28e$ was necessary to reproduce the \beup with $A=106-114$, while assuming $e_{\nu}=1.0e$ vastly 
%underestimated the experimental values. For \ts{104}Sn, a \beup value of 0.14 \eb is predicted, in good agreement with our result
%and implying a persistent large \ts{100}Sn core polarization. 
In an alternative approach that included the neutron $gds$ and $0h_{11/2}$ orbitals as model space and single-particle energies
fitted to experimental data~\cite{qi:2012:PRC}, isospin-dependent effective charges as proposed by Bohr and 
Mottelson~\cite{bohrmottelson:1975} were introduced into the calculations~\cite{baeck:2013:PRC}. The neglect of proton excitations was 
compensated by normalizing the effective charges to $e_{\nu} = 1.0e$ in the middle of the shell for \ts{116}Sn resulting 
in  $e_{\nu} > 1.0e$ ($e_{\nu} < 1.0e$) in the lower (upper) half of the shell. Indeed, a good overall agreement is observed for 
very neutron and proton rich nuclei, as shown in Fig.~\ref{fig:systematics} by LSSM\ts{d}. However, the collectivity increase
on the proton-rich side commences later than observed in experiments~\cite{doornenbal:2008:PRC,kumar:2010:PRC,jungclaus:2011:PLB} and 
the large effective charges are also coincident with the proton gap minimum around 
\ts{108}Sn~(see, e.g., Fig.~4 of Ref.~\cite{doornenbal:2008:PRC}). 
 
In combination with the neglect of proton excitations,
this effective charge adjustment can therefore only be regarded as an interim solution until sufficient computing power
becomes available. The importance of proton excitations across the $Z=50$ shell for a correct description of the \beup pattern
can be easily inferred from the difference in ``matter'' deformation lengths obtained from the carbon target and 
the charge deformation lengths obtained from the \beup values. With $\delta_{c} = (4 \pi / 3eZ R_{0}) B(E2)\!\!\uparrow^{1/2}$  
and $R_{0} = 1.2 A^{1/3}$~fm, the latter are $\delta_{c} = 0.59$(9) and 0.71(2)~fm for \ts{104,112}Sn and hence about 50~\% larger 
than the values of $\delta = 0.38(2)$ and 0.45(2)~fm obtained with the carbon target. 

% Summary 
In summary, a \beup value of \betwosn \eb was measured for \ts{104}Sn in intermediate-energy Coulomb 
excitation. The drop in excitation strength is much smoother than obtained in \cite{guastalla:2013:PRL} and 
cannot be reproduced by present LSSM calculations using standard effective charges as well as proton
and neutron excitation across the $N=Z=50$ shell. 
%Therefore, also the formation of a \beup strength minimum for
%\ts{102}Sn predicted by the LSSM\ts{c} calculations remains an open question.
%Correspondence between \beup values and proton gap deduced from the AME2012 
%suggests that future \beup calculations should aim to simultaneously reproduce the
%binding energies including neighboring isotones. 
Moreover, it was demonstrated that given the significant scattering angle resolution and angular
straggling at energies well above 100 \mevu, nuclear excitation should be explicitly taken into
account in \beup determinations rather than suppressed in an inaccurate angular cut. Coulomb feeding from
higher lying states cannot be neglected but can easily be determined from known \beup
values. A simple scaling of measured cross-sections for the \ts{104,112}Sn pair would have led to a 10~\%
lower \beup assignment for \ts{104}Sn. Therefore, we suggest that future absolute cross-section measurements at 
energies well above 100 \mevu include calibration runs of nuclei with known \beup values on a high $Z$ target and 
nuclear excitation on a low $Z$ target. Such an approach allows for the use of very thick reaction targets and
thus gives access to more exotic nuclei. 
%For experiments performed at the RIBF, we do not recommend to reduce the secondary beam energies 
%due to the angular acceptance of the ZeroDegree spectrometer, the reduced beam intensities, and
%increasing charge state distributions. 
% (comment?)  
 
\begin{acknowledgments}

We would like to thank the RIKEN Nishina Center Accelerator Division for providing the high
\ts{124}Xe primary beam intensity and the BigRIPS team for preparing secondary beams
with high purities for the isotopes of interest. We acknowledge financial support from the 
Spanish Ministerio de Ciencia e Innovaci\'on under contracts FPA2009-13377-C02-02 and
FPA2011-29854-C04-01, the OTKA under contract number K100835, and the European Research 
Council through the ERC Starting Grant MINOS-258567. 
\end{acknowledgments}

% 
% references
% 
\bibliography{local}   % most articles can be referenced as \cite{last-name:year:journal}

\begin{thebibliography}{36}
\expandafter\ifx\csname natexlab\endcsname\relax\def\natexlab#1{#1}\fi
\expandafter\ifx\csname bibnamefont\endcsname\relax
  \def\bibnamefont#1{#1}\fi
\expandafter\ifx\csname bibfnamefont\endcsname\relax
  \def\bibfnamefont#1{#1}\fi
\expandafter\ifx\csname citenamefont\endcsname\relax
  \def\citenamefont#1{#1}\fi
\expandafter\ifx\csname url\endcsname\relax
  \def\url#1{\texttt{#1}}\fi
\expandafter\ifx\csname urlprefix\endcsname\relax\def\urlprefix{URL }\fi
\providecommand{\bibinfo}[2]{#2}
\providecommand{\eprint}[2][]{\url{#2}}

\bibitem[{\citenamefont{Wang et~al.}(2012)\citenamefont{Wang, Audi, Wapstra,
  Kondev, MacCormick, Xu, and Pfeiffer}}]{wang:2012:CPC}
\bibinfo{author}{\bibfnamefont{M.}~\bibnamefont{Wang}},
  \bibinfo{author}{\bibfnamefont{G.}~\bibnamefont{Audi}},
  \bibinfo{author}{\bibfnamefont{A.}~\bibnamefont{Wapstra}},
  \bibinfo{author}{\bibfnamefont{F.}~\bibnamefont{Kondev}},
  \bibinfo{author}{\bibfnamefont{M.}~\bibnamefont{MacCormick}},
  \bibinfo{author}{\bibfnamefont{X.}~\bibnamefont{Xu}}, \bibnamefont{and}
  \bibinfo{author}{\bibfnamefont{B.}~\bibnamefont{Pfeiffer}},
  \bibinfo{journal}{Chin. Phys. C} \textbf{\bibinfo{volume}{36}},
  \bibinfo{pages}{1603} (\bibinfo{year}{2012}).

\bibitem[{\citenamefont{Hinke et~al.}(2012)}]{hinke:2012:NATURE}
\bibinfo{author}{\bibfnamefont{C.}~\bibnamefont{Hinke}} \bibnamefont{et~al.},
  \bibinfo{journal}{Nature} \textbf{\bibinfo{volume}{486}},
  \bibinfo{pages}{341} (\bibinfo{year}{2012}).

\bibitem[{\citenamefont{Lipoglavsek et~al.}(1996)}]{lipoglavsek:1996:ZPA}
\bibinfo{author}{\bibfnamefont{M.}~\bibnamefont{Lipoglavsek}}
  \bibnamefont{et~al.}, \bibinfo{journal}{Z. Phys. A}
  \textbf{\bibinfo{volume}{356}}, \bibinfo{pages}{239} (\bibinfo{year}{1996}).

\bibitem[{\citenamefont{Raman et~al.}(2001)\citenamefont{Raman, Jr., and
  Tikkanen}}]{raman:2001:ADNDT}
\bibinfo{author}{\bibfnamefont{S.}~\bibnamefont{Raman}},
  \bibinfo{author}{\bibfnamefont{C.~W.~N.} \bibnamefont{Jr.}},
  \bibnamefont{and} \bibinfo{author}{\bibfnamefont{P.}~\bibnamefont{Tikkanen}},
  \bibinfo{journal}{Atom. Data and Nucl. Data Tab.}
  \textbf{\bibinfo{volume}{78}}, \bibinfo{pages}{1} (\bibinfo{year}{2001}).

\bibitem[{\citenamefont{Casten}(2001)}]{casten:2001}
\bibinfo{author}{\bibfnamefont{R.}~\bibnamefont{Casten}},
  \emph{\bibinfo{title}{Nuclear Structure from a Simple Perspective}}
  (\bibinfo{publisher}{Oxford University Press}, \bibinfo{year}{2001}).

\bibitem[{\citenamefont{Morales et~al.}(2011)\citenamefont{Morales, Isacker,
  and Talmi}}]{morales:2011:PLB}
\bibinfo{author}{\bibfnamefont{I.}~\bibnamefont{Morales}},
  \bibinfo{author}{\bibfnamefont{P.~V.} \bibnamefont{Isacker}},
  \bibnamefont{and} \bibinfo{author}{\bibfnamefont{I.}~\bibnamefont{Talmi}},
  \bibinfo{journal}{Phys. Lett. B} \textbf{\bibinfo{volume}{703}},
  \bibinfo{pages}{606} (\bibinfo{year}{2011}).

\bibitem[{\citenamefont{Radford et~al.}(2004)}]{radford:2004:NPA}
\bibinfo{author}{\bibfnamefont{D.}~\bibnamefont{Radford}} \bibnamefont{et~al.},
  \bibinfo{journal}{Nucl. Phys. A} \textbf{\bibinfo{volume}{746}},
  \bibinfo{pages}{83c} (\bibinfo{year}{2004}).

\bibitem[{\citenamefont{Allmond et~al.}(2011)}]{allmond:2011:PRC}
\bibinfo{author}{\bibfnamefont{J.}~\bibnamefont{Allmond}} \bibnamefont{et~al.},
  \bibinfo{journal}{Phys. Rev. C} \textbf{\bibinfo{volume}{84}},
  \bibinfo{pages}{061303(R)} (\bibinfo{year}{2011}).

\bibitem[{\citenamefont{Banu et~al.}(2005)}]{banu:2005:PRC}
\bibinfo{author}{\bibfnamefont{A.}~\bibnamefont{Banu}} \bibnamefont{et~al.},
  \bibinfo{journal}{Phys. Rev. C} \textbf{\bibinfo{volume}{72}},
  \bibinfo{pages}{061305} (\bibinfo{year}{2005}).

\bibitem[{\citenamefont{Ekstr\"om et~al.}(2008)}]{ekstrom:2008:PRL}
\bibinfo{author}{\bibfnamefont{A.}~\bibnamefont{Ekstr\"om}}
  \bibnamefont{et~al.}, \bibinfo{journal}{Phys. Rev. Lett.}
  \textbf{\bibinfo{volume}{101}}, \bibinfo{pages}{012502}
  (\bibinfo{year}{2008}).

\bibitem[{\citenamefont{Cederk\"all et~al.}(2007)}]{cederkall:2007:PRL}
\bibinfo{author}{\bibfnamefont{J.}~\bibnamefont{Cederk\"all}}
  \bibnamefont{et~al.}, \bibinfo{journal}{Phys. Rev. Lett.}
  \textbf{\bibinfo{volume}{98}}, \bibinfo{pages}{172501}
  (\bibinfo{year}{2007}).

\bibitem[{\citenamefont{Vaman et~al.}(2007)}]{vaman:2007:PRL}
\bibinfo{author}{\bibfnamefont{C.}~\bibnamefont{Vaman}} \bibnamefont{et~al.},
  \bibinfo{journal}{Phys. Rev. Lett.} \textbf{\bibinfo{volume}{99}},
  \bibinfo{pages}{162501} (\bibinfo{year}{2007}).

\bibitem[{\citenamefont{Doornenbal et~al.}(2008)}]{doornenbal:2008:PRC}
\bibinfo{author}{\bibfnamefont{P.}~\bibnamefont{Doornenbal}}
  \bibnamefont{et~al.}, \bibinfo{journal}{Phys. Rev. C}
  \textbf{\bibinfo{volume}{78}}, \bibinfo{pages}{031303}
  (\bibinfo{year}{2008}).

\bibitem[{\citenamefont{Kumar et~al.}(2010)}]{kumar:2010:PRC}
\bibinfo{author}{\bibfnamefont{R.}~\bibnamefont{Kumar}} \bibnamefont{et~al.},
  \bibinfo{journal}{Phsys. Rev. C} \textbf{\bibinfo{volume}{81}},
  \bibinfo{pages}{024306} (\bibinfo{year}{2010}).

\bibitem[{\citenamefont{Jungclaus et~al.}(2011)}]{jungclaus:2011:PLB}
\bibinfo{author}{\bibfnamefont{A.}~\bibnamefont{Jungclaus}}
  \bibnamefont{et~al.}, \bibinfo{journal}{Phys. Lett. B}
  \textbf{\bibinfo{volume}{110}}, \bibinfo{pages}{695} (\bibinfo{year}{2011}).

\bibitem[{\citenamefont{Guastalla et~al.}(2013)}]{guastalla:2013:PRL}
\bibinfo{author}{\bibfnamefont{G.}~\bibnamefont{Guastalla}}
  \bibnamefont{et~al.}, \bibinfo{journal}{Phys. Rev. Lett.}
  \textbf{\bibinfo{volume}{110}}, \bibinfo{pages}{172501}
  (\bibinfo{year}{2013}).

\bibitem[{\citenamefont{B\"urger et~al.}(2005)}]{burger:2005:PLB}
\bibinfo{author}{\bibfnamefont{A.}~\bibnamefont{B\"urger}}
  \bibnamefont{et~al.}, \bibinfo{journal}{Phys. Lett. B}
  \textbf{\bibinfo{volume}{622}}, \bibinfo{pages}{29} (\bibinfo{year}{2005}).

\bibitem[{\citenamefont{Saito et~al.}(2008)}]{saito:2008:PLB}
\bibinfo{author}{\bibfnamefont{T.}~\bibnamefont{Saito}} \bibnamefont{et~al.},
  \bibinfo{journal}{Phys. Lett. B} \textbf{\bibinfo{volume}{669}},
  \bibinfo{pages}{19} (\bibinfo{year}{2008}).

\bibitem[{\citenamefont{Kubo et~al.}(2012)}]{kubo:2012:PTEP}
\bibinfo{author}{\bibfnamefont{T.}~\bibnamefont{Kubo}} \bibnamefont{et~al.},
  \bibinfo{journal}{Prog. Theor. Exp. Phys.} \textbf{\bibinfo{volume}{2012}},
  \bibinfo{pages}{03C003} (\bibinfo{year}{2012}).

\bibitem[{\citenamefont{Kumagai et~al.}(2001)\citenamefont{Kumagai, Ozawa,
  Fukuda, S\"ummerer, and Tanihata}}]{kumagai:2001:NIMA}
\bibinfo{author}{\bibfnamefont{H.}~\bibnamefont{Kumagai}},
  \bibinfo{author}{\bibfnamefont{A.}~\bibnamefont{Ozawa}},
  \bibinfo{author}{\bibfnamefont{N.}~\bibnamefont{Fukuda}},
  \bibinfo{author}{\bibfnamefont{K.}~\bibnamefont{S\"ummerer}},
  \bibnamefont{and} \bibinfo{author}{\bibfnamefont{I.}~\bibnamefont{Tanihata}},
  \bibinfo{journal}{Nucl. Instrum. and Meth. A} \textbf{\bibinfo{volume}{470}},
  \bibinfo{pages}{562} (\bibinfo{year}{2001}).

\bibitem[{ati()}]{atima}
\emph{\bibinfo{title}{Atima}},
  \bibinfo{howpublished}{\url{http://http://web-docs.gsi.de/~weick/atima/}}.

\bibitem[{\citenamefont{Wollersheim et~al.}(2005)}]{wollersheim:2005:nima}
\bibinfo{author}{\bibfnamefont{H.}~\bibnamefont{Wollersheim}}
  \bibnamefont{et~al.}, \bibinfo{journal}{Nucl. Instrum. and Meth. A}
  \textbf{\bibinfo{volume}{537}}, \bibinfo{pages}{637} (\bibinfo{year}{2005}).

\bibitem[{\citenamefont{Takeuchi et~al.}(2003)}]{takeuchi:2003}
\bibinfo{author}{\bibfnamefont{S.}~\bibnamefont{Takeuchi}}
  \bibnamefont{et~al.}, in \emph{\bibinfo{booktitle}{RIKEN Accelerator Progress
  Report}} (\bibinfo{publisher}{RIKEN}, \bibinfo{year}{2003}),
  vol.~\bibinfo{volume}{36}, p. \bibinfo{pages}{148}.

\bibitem[{\citenamefont{Agostinelli et~al.}(2003)}]{agostinelli:2003:nima}
\bibinfo{author}{\bibfnamefont{S.}~\bibnamefont{Agostinelli}}
  \bibnamefont{et~al.}, \bibinfo{journal}{Nucl. Instr. Meth. A}
  \textbf{\bibinfo{volume}{506}}, \bibinfo{pages}{250} (\bibinfo{year}{2003}).

\bibitem[{\citenamefont{Bertulani et~al.}(2003)\citenamefont{Bertulani,
  Campbell, and Glasmacher}}]{bertulani:2003:CPC}
\bibinfo{author}{\bibfnamefont{C.~A.} \bibnamefont{Bertulani}},
  \bibinfo{author}{\bibfnamefont{C.~M.} \bibnamefont{Campbell}},
  \bibnamefont{and}
  \bibinfo{author}{\bibfnamefont{T.}~\bibnamefont{Glasmacher}},
  \bibinfo{journal}{Comp. Phys. Com.} \textbf{\bibinfo{volume}{152}},
  \bibinfo{pages}{317} (\bibinfo{year}{2003}).

\bibitem[{\citenamefont{Furumoto et~al.}(2012)\citenamefont{Furumoto,
  W.Horiuchi, M.Takashina, Y.Yamamoto, and Y.Sakuragi}}]{furumoto:2012:PRC}
\bibinfo{author}{\bibfnamefont{T.}~\bibnamefont{Furumoto}},
  \bibinfo{author}{\bibnamefont{W.Horiuchi}},
  \bibinfo{author}{\bibnamefont{M.Takashina}},
  \bibinfo{author}{\bibnamefont{Y.Yamamoto}}, \bibnamefont{and}
  \bibinfo{author}{\bibnamefont{Y.Sakuragi}}, \bibinfo{journal}{Phys. Rev. C}
  \textbf{\bibinfo{volume}{85}}, \bibinfo{pages}{044607}
  (\bibinfo{year}{2012}).

\bibitem[{\citenamefont{Furumoto et~al.}(2008)\citenamefont{Furumoto, Sakuragi,
  and Yamamoto}}]{furumoto:2008:PRC}
\bibinfo{author}{\bibfnamefont{T.}~\bibnamefont{Furumoto}},
  \bibinfo{author}{\bibfnamefont{Y.}~\bibnamefont{Sakuragi}}, \bibnamefont{and}
  \bibinfo{author}{\bibfnamefont{Y.}~\bibnamefont{Yamamoto}},
  \bibinfo{journal}{Phys. Rev. C} \textbf{\bibinfo{volume}{78}},
  \bibinfo{pages}{044610} (\bibinfo{year}{2008}).

\bibitem[{\citenamefont{Furumoto et~al.}(2009)\citenamefont{Furumoto, Sakuragi,
  and Yamamoto}}]{furumoto:2009:PRC}
\bibinfo{author}{\bibfnamefont{T.}~\bibnamefont{Furumoto}},
  \bibinfo{author}{\bibfnamefont{Y.}~\bibnamefont{Sakuragi}}, \bibnamefont{and}
  \bibinfo{author}{\bibfnamefont{Y.}~\bibnamefont{Yamamoto}},
  \bibinfo{journal}{Phys. Rev. C} \textbf{\bibinfo{volume}{80}},
  \bibinfo{pages}{044614} (\bibinfo{year}{2009}).

\bibitem[{\citenamefont{Chamon et~al.}(2002)}]{chamon:2002:PRC}
\bibinfo{author}{\bibfnamefont{L.}~\bibnamefont{Chamon}} \bibnamefont{et~al.},
  \bibinfo{journal}{Phys. Rev. C} \textbf{\bibinfo{volume}{66}},
  \bibinfo{pages}{014610} (\bibinfo{year}{2002}).

\bibitem[{\citenamefont{Jonsson et~al.}(1981)\citenamefont{Jonsson, Backlin,
  Kantele, Julin, Luontama, and Passoja}}]{jonsson:1981:NPA}
\bibinfo{author}{\bibfnamefont{N.}~\bibnamefont{Jonsson}},
  \bibinfo{author}{\bibfnamefont{A.}~\bibnamefont{Backlin}},
  \bibinfo{author}{\bibfnamefont{J.}~\bibnamefont{Kantele}},
  \bibinfo{author}{\bibfnamefont{R.}~\bibnamefont{Julin}},
  \bibinfo{author}{\bibfnamefont{M.}~\bibnamefont{Luontama}}, \bibnamefont{and}
  \bibinfo{author}{\bibfnamefont{A.}~\bibnamefont{Passoja}},
  \bibinfo{journal}{Nuclear Physics A} \textbf{\bibinfo{volume}{371}},
  \bibinfo{pages}{333 } (\bibinfo{year}{1981}).

\bibitem[{\citenamefont{Bryssinck et~al.}(2000)}]{bryssinck:2000:PRC}
\bibinfo{author}{\bibfnamefont{A.}~\bibnamefont{Bryssinck}}
  \bibnamefont{et~al.}, \bibinfo{journal}{Phys. Rev. C}
  \textbf{\bibinfo{volume}{61}}, \bibinfo{pages}{024309}
  (\bibinfo{year}{2000}).

\bibitem[{\citenamefont{Jiang et~al.}(2012)\citenamefont{Jiang, Lei, Fu, Zhao,
  and Arima}}]{jiang:2012:PRC}
\bibinfo{author}{\bibfnamefont{H.}~\bibnamefont{Jiang}},
  \bibinfo{author}{\bibfnamefont{Y.}~\bibnamefont{Lei}},
  \bibinfo{author}{\bibfnamefont{G.}~\bibnamefont{Fu}},
  \bibinfo{author}{\bibfnamefont{Y.}~\bibnamefont{Zhao}}, \bibnamefont{and}
  \bibinfo{author}{\bibfnamefont{A.}~\bibnamefont{Arima}},
  \bibinfo{journal}{Phys. Rev. C} \textbf{\bibinfo{volume}{86}},
  \bibinfo{pages}{054304} (\bibinfo{year}{2012}).

\bibitem[{\citenamefont{B\"ack et~al.}(2013)\citenamefont{B\"ack, Qi,
  Cederwall, Liotta, Moradi, Johnson, Wyss, and Wadsworth}}]{baeck:2013:PRC}
\bibinfo{author}{\bibfnamefont{T.}~\bibnamefont{B\"ack}},
  \bibinfo{author}{\bibfnamefont{C.}~\bibnamefont{Qi}},
  \bibinfo{author}{\bibfnamefont{B.}~\bibnamefont{Cederwall}},
  \bibinfo{author}{\bibfnamefont{R.}~\bibnamefont{Liotta}},
  \bibinfo{author}{\bibfnamefont{F.~G.} \bibnamefont{Moradi}},
  \bibinfo{author}{\bibfnamefont{A.}~\bibnamefont{Johnson}},
  \bibinfo{author}{\bibfnamefont{R.}~\bibnamefont{Wyss}}, \bibnamefont{and}
  \bibinfo{author}{\bibfnamefont{R.}~\bibnamefont{Wadsworth}},
  \bibinfo{journal}{Phys. Rev. C} \textbf{\bibinfo{volume}{87}},
  \bibinfo{pages}{031306(R)} (\bibinfo{year}{2013}).

\bibitem[{\citenamefont{Machleidt}(2001)}]{machleidt:2001:PRC}
\bibinfo{author}{\bibfnamefont{R.}~\bibnamefont{Machleidt}},
  \bibinfo{journal}{Phys. Rev. C} \textbf{\bibinfo{volume}{63}},
  \bibinfo{pages}{024001} (\bibinfo{year}{2001}).

\bibitem[{\citenamefont{Qi and Xu}(2012)}]{qi:2012:PRC}
\bibinfo{author}{\bibfnamefont{C.}~\bibnamefont{Qi}} \bibnamefont{and}
  \bibinfo{author}{\bibfnamefont{Z.}~\bibnamefont{Xu}}, \bibinfo{journal}{Phys.
  Rev. C} \textbf{\bibinfo{volume}{86}}, \bibinfo{pages}{044323}
  (\bibinfo{year}{2012}).

\bibitem[{\citenamefont{Bohr and B.Mottelson}(1975)}]{bohrmottelson:1975}
\bibinfo{author}{\bibfnamefont{A.}~\bibnamefont{Bohr}} \bibnamefont{and}
  \bibinfo{author}{\bibnamefont{B.Mottelson}}, \emph{\bibinfo{title}{Nuclear
  Structure, Vol. II}} (\bibinfo{publisher}{Benjamin, New York},
  \bibinfo{year}{1975}).

\end{thebibliography}
\end{document}